\documentclass[conference]{IEEEtran}
\IEEEoverridecommandlockouts

\usepackage{cite}
\usepackage{amsmath,amssymb,amsfonts}
\usepackage{algorithmic}
\usepackage{textcomp}
\usepackage{xcolor}

\usepackage{graphicx}
\usepackage{enumitem}
\usepackage{multirow}
\usepackage{pifont}
\usepackage{subfigure}
\usepackage{mathtools}

\usepackage{hyperref}       
\usepackage{url}            
\usepackage{booktabs}       
\usepackage{nicefrac}       
\usepackage{microtype}      

\usepackage{multirow}   
\usepackage{graphicx}
\usepackage{colortbl}
\usepackage{multicol}
\usepackage{array}

\usepackage[ruled,vlined]{algorithm2e}
\usepackage{multirow}

\def\BibTeX{{\rm B\kern-.05em{\sc i\kern-.025em b}\kern-.08em
    T\kern-.1667em\lower.7ex\hbox{E}\kern-.125emX}}
\begin{document}

\title{\LARGE \bf Primer: Fast Private Transformer Inference on Encrypted Data }

\author{
\begin{tabular}{c c c}
Mengxin Zheng  & Qian Lou   & Lei Jiang  \\
\multicolumn{1}{c}{Indiana University Bloomington} & \multicolumn{1}{c}{ University of Central Florida} & \multicolumn{1}{c}{Indiana University Bloomington}\\
\multicolumn{1}{c}{ zhengme@iu.edu} & \multicolumn{1}{c}{ qian.lou@ucf.edu} &
\multicolumn{1}{c}{ jiang60@iu.edu}\\
\end{tabular}
}
\maketitle

\begin{abstract}


It is increasingly important to enable privacy-preserving inference for cloud services based on Transformers. Post-quantum cryptographic techniques, e.g., fully homomorphic encryption (FHE), and multi-party computation (MPC), are popular methods to support private Transformer inference. However, existing works still suffer from prohibitively computational and communicational overhead. In this work,  we present, Primer, to enable a fast and accurate Transformer over encrypted data for natural language processing tasks. In particular, Primer is constructed by a hybrid cryptographic protocol optimized for attention-based Transformer models, as well as techniques including computation merge  and tokens-first ciphertext packing. Comprehensive experiments on encrypted language modeling show that Primer achieves state-of-the-art accuracy and reduces the inference latency by $90.6\% \sim 97.5\%$ over previous methods.

\end{abstract}


\section{Introduction}
\label{s:intro}

Transformer-based, or more broadly attention-based, models show superior performance over previous methods, becoming increasingly popular in natural language processing (NLP) applications~\cite{Devlin:ACL2019:BERT}. For example, BERT obtains new state-of-the-art results on eleven NLP tasks, including pushing the GLUE score to $80.5\%$ ($7.7\%$  absolute improvement), and even proves superior performance compared to human results on the challenging sentence classification tasks. 
Server-based Transformer service is an effective way for clients to run their computationally expensive and memory-intensive NLP tasks on powerful cloud servers. During a server-based Transformer service, cloud servers require access to clients' language data, thus introducing potential privacy risks. Therefore, to be able to utilize this technology, it is urgently needed to safeguard the confidentiality of users'  biomedical, financial, and other sensitive data that are submitted to servers. 

\begin{figure}[t!]
\centering
\includegraphics[width=3.3in]{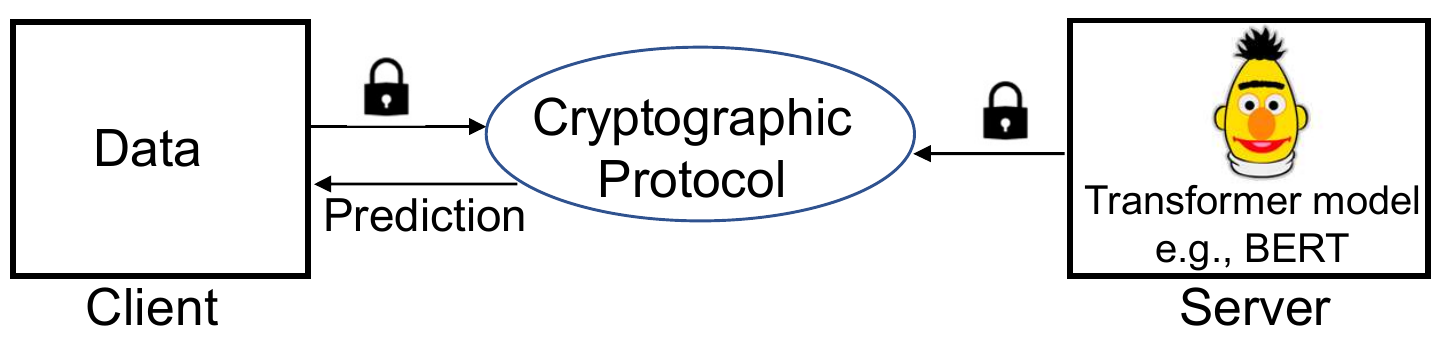}
\vspace{-0.1in}
\caption{Overview of private Transformer inference based on cryptographic protocol, e.g., FHE and MPC. The lock represents that data is encrypted. }
\vspace{-0.2in}
\label{f:overview}
\end{figure}

Post-quantum cryptographic protocols, e.g., FHE~\cite{FHE,Lou:NIPS2019} and MPC~\cite{Mihir:Justgarble} are popular methods to enable provably confidential computation on encrypted data. We use Figure~\ref{f:overview} to show the overview of private transformer inference, where the client receives cloud services based on Transformer models by only uploading encrypted data generated by cryptographic protocols such as FHE or MPC. This Transformer inference is provably privacy-preserving since data is not revealed to other parties~\cite{li2022mpcformer,chen:ACL2022}.  However, existing  works for private Transformer inference based on FHE, e.g., THE-X~\cite{chen:ACL2022}, suffer from enormous latency. For example, THE-X takes more than 3 orders of magnitudes latency than regular Transformer inference. And polynomial approximation of activation in THE-X significantly reduces accuracy, e.g., $<77\%$ GLUE score ($\sim 8\%$ absolute accuracy decrease).  



We identify several challenges to design private Transformer inference, such as the large one-hot word embeddings, complex attention, frequent $SoftMax$, and very deep blocks. Specifically, BERT~\cite{Devlin:ACL2019:BERT} uses WordPiece embeddings~\cite{google:2016:wordpiece} with $30522$ token vocabulary and $768$ embedding dimensions so that $n$ tokens require $n$ times of $30522\times768$ matrix-vector multiplication. Directly applying existing techniques to design privacy-preserving embeddings suffers from enormous latency overhead. In addition,  we identify that the attention scheme in Transformer models requires massive ciphertext-ciphertext multiplications that cannot directly be implemented by previous methods that are optimized for ciphertext-plaintext multiplications.  Moreover, deeper Transformer architecture adds expensive FHE rotations and communicational interactions. 

In this work, we present a fast and accurate Transformer inference method, denoted by Primer, over encrypted data. We propose several techniques to construct Primer. In particular, a hybrid cryptographic protocol is proposed to construct a private Transformer, where FHE is used for polynomial operations and MPC is for non-polynomial operations. We call our Primer with this protocol Primer-base. Primer-base is accurate since it removes the polynomial approximation in previous works based on FHE. To reduce the online time of Primer-base, we propose a new hybrid protocol, denoted by HGS,  to pre-process most FHE operations. Offloading computations into the offline phase from the online phase is important since offline computations can be computed in advance before inference.  We further propose FHGS, denoted by Primer-F to improve the compatibility of HGS on attention computations in Transformer models. Other techniques including computation merge (combined FHGS) and tokens-first packing are presented to further reduce the inference latency.  
Comprehensive experiments on encrypted language modeling show that Primer achieves state-of-the-art accuracy and reduces the inference latency by $90.6\% \sim 97.5\%$ over previous methods.

\section{Background and Motivation}
\textbf{Threat Model.} 
We use Figure~\ref{f:overview} to show the overview of our threat model, where servers and clients are semi-honest, e.g.,  a semi-honest cloud server that attempts to infer clients' input information but follows the cryptographic protocol. Our threat model follows previous work THE-X ~\cite{chen:ACL2022} and Gazelle~\cite{GAZELLE:USENIX18}.  The security level of our method is 128 bits for a fair comparison.

\textbf{Transformer-based Models for NLP Tasks.}
Transformer-based models~\cite{Vaswani:NIPS2017:Attention}  achieve state-of-the-art performance in many NLP tasks. 
A Transformer architecture mainly includes embeddings, stacked encoders, and decoders using Multi-Head Self-Attention (MHSA) and point-wise, fully connected (FC) layers. A model with only encoders, e.g. BERT~\cite{Devlin:ACL2019:BERT}, can be used in discriminative NLP tasks including classification and regression, etc. Meanwhile, a model with decoders, e.g. GPT-2~\cite{radford:2019:GPT2}, works for generative NLP tasks including Language Modeling (LM) and machine translation.   In particular, embeddings include word embedding and positional embedding. Word embedding converts the input tokens and output tokens (each token is a one-hot vector with a length of $d_{oh}$) to vectors of dimension $d_{emb}$ by a linear projection. Positional embedding ensures the Transformer model has the sequence order information by adding "positional encodings" $\lambda$ to the previous word embedding. The embedded representations are fed into MHSA. In MHSA, embedded representations are firstly converted into three categories, key $X_K$, query $X_Q$, and value $X_V$, by linear projections with key weight $W_K$, query weight $W_Q$, and value weight $W_V$, respectively. Then, the output of MHSA is calculated as a weighted sum of the values $X_V$ by $Attention(X_Q, X_K, X_V)=SoftMax(\frac{X_QX_K^T}{\sqrt{n}})X_V$, where $SoftMax(\frac{X_QX_K^T}{\sqrt{n}})$ is the weight assigned to each value, and $n$ is token numbers. Instead of only computing the attention once, the multi-head mechanism computes attention multiple times in parallel, and these multiple attentions are simply concatenated and linearly transformed into the expected dimensions as $MultiHead(X_Q,X_K,X_V)=[head_1,..., head_H]W_O$, where $head_i=Attention(X_QW_Q^i, X_KW_K^i, X_VW_V^i)$, $W_O$ is a linear projection weight matrix. 

\textbf{Interactive hybrid cryptographic protocol.}
FHE~\cite{FHE} is an encryption method that enables one to perform computations on encrypted data without decryption. Garbled Circuit (GC)~\cite{Mihir:Justgarble,feng:2020:cryptogru} and Secret Sharing (SS)~\cite{GMW:1987:SS} are two paramount methods of multi-party secure computations. An Interactive hybrid cryptographic protocol~\cite{GAZELLE:USENIX18} is proposed to combine the advantages of FHE, GC, and SS. In particular, FHE has superior performance over GC on linear operations, e.g., matrix-vector multiplication. This is because FHE with a ciphertext packing technique supports efficient operations in a SIMD (single instruction multiple data) manner.   Therefore, FHE is used to support private linear operations where a client encrypts input and sends it to the server, and the server returns encrypted output to the client that decrypts the received output.  In the state-of-the-art mixed protocols~\cite{GAZELLE:USENIX18,Lou:ICLR2021:safenet,Lou:NIPS20:AutoPrivacy}, GC shows superior performance over HE in non-linear operations such as activation functions. And SS is used to combine GC and HE in the mixed protocol. Inspired by Beaver's Triple~\cite{beaver1995}, FHE can be used to efficiently perform multiplications on two additive secret shares.  In this work, we use this interactive hybrid method to construct Primer-base, which is a starting point for our optimization techniques.

\begin{figure}[t!]
\centering
\includegraphics[width=3.3in]{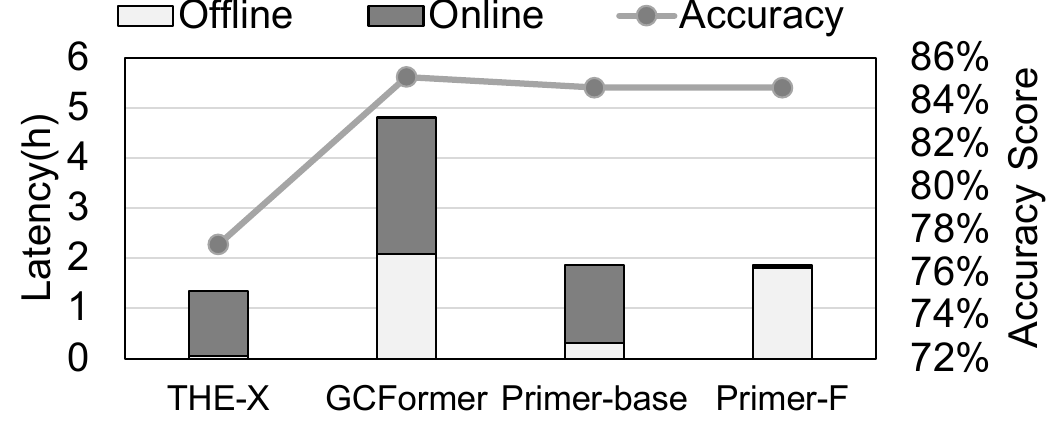}
\caption{The latency and accuracy comparisons of prior works, e.g., THE-X and GCFormer, and our work Primer-base and Primer-F on MNLI-m dataset with BERT-base.}
\vspace{-0.2in}
\label{f:motivation-dac}
\end{figure}

\textbf{Motivation.} 
As Figure~\ref{f:motivation-dac} shows, prior works like THE-X~\cite{chen:ACL2022} using only FHE for private inference suffer from low accuracy and enormous online latency due to polynomial approximation and expensive FHE operations.  We use prior GC-based work~\cite{Deepsecure:dac18} to implement a GCFormer (we convert the Transformer model into a circuit based on binary gates so that GC~\cite{Mihir:Justgarble} can implement it). GCFormer achieves an accurate performance, i.e., 85.1\% accuracy, but it takes a larger latency than THE-X. Thus, the FHE-based method or GC-based method  cannot achieve a low-latency and accurate private Transformer inference. Instead, we follow the interactive and hybrid cryptographic protocol~\cite{GAZELLE:USENIX18} and construct our Primer-base by using GC for non-polynomial operations, FHE for polynomial operations, and SS for secure communication between multiple parties.  Primer-base significantly improves the accuracy of THE-X, e.g., 7.3\% accuracy increase, and reduces the latency of GCFormer.  However, Primer-base still suffers from enormous online latency. This motivates us to propose techniques like the FHGS protocol, denoted by Primer-F, to offload the online computation to the offline phase where computations can be computed before inference.  Considering Primer-F still has a large total latency, we have motivations to propose techniques including computation merge, i.e., combined FHGS, and tokens-first ciphertext packing techniques. More details about Primer and related techniques are introduced in the following section~\ref{s:Primer}.

\section{Primer}
\label{s:Primer}

\subsection{Primer-base Construction}
\label{s:Primer-base}

\begin{figure}[t!]
\centering
\includegraphics[width=3.5in]{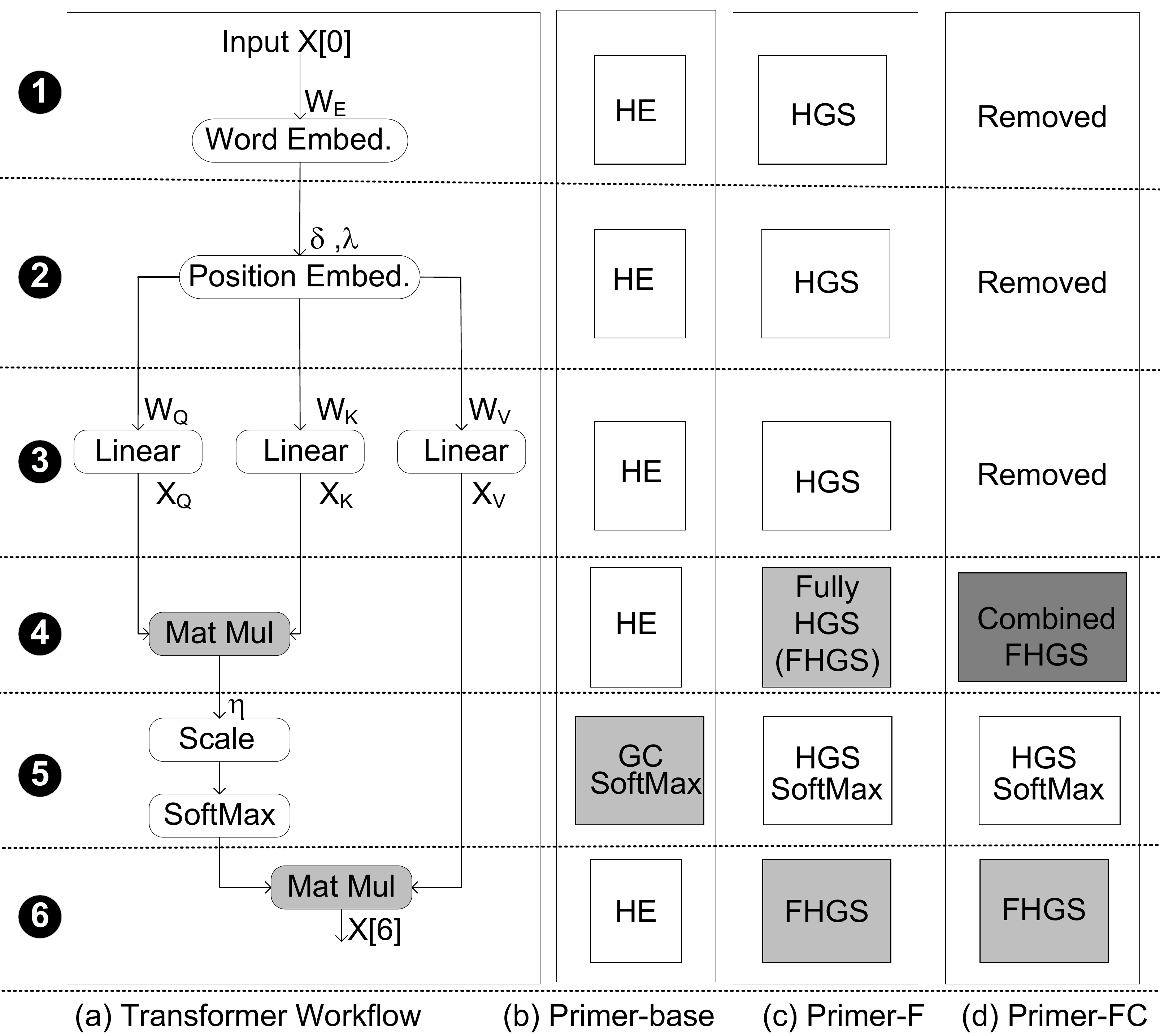}
\caption{Private transformer block inference under various Primer protocols.}
\vspace{-0.2in}
\label{f:flow}
\end{figure}

As Figure~\ref{f:flow}(a) shows, a Transformer-based model involves computations of \ding{182}-\ding{183} embeddings, \ding{184} derivation of $X_Q$, $X_K$ and $X_V$, \ding{185} scaled dot-product of $X_Q$ and $X_K^T$, \ding{186} $SoftMax$, \ding{187} attention values, and the other linear operations like Fully Connected (FC) computations. 
Embedding output $X[1]$ is computed by $X[0]\times W_E \times \delta +\lambda$, where $X[0]$, $W_E$, $\delta$, and $\lambda$ are the input matrix, embedding weight matrix, positional coefficients, and positional biases, respectively. The embedding output $X[1]$ is multiplied with query weight $W_Q$, key weight $W_K$, and value weight $W_V$ to generate query $X_Q$, key $X_K$, and value $X_V$. Multi-Head Self-Attention requires multiple computations of $X_Q$, key $X_K$, and value $X_V$ with various weight matrices in parallel.  Then a Transformer needs to compute the dot products between the query $X_Q$ and all keys $X_K$, divide each by $\eta =\sqrt{n}$, apply a $SoftMax$ function to obtain the weights on the values, and multiply the attention weights with the value $X_V$.  Figure~\ref{f:flow}(b) illustrates how to construct a basic private Transformer, i.e., Primer-base, using the prior interactive hybrid cryptographic protocol, i.e., FHE (we denote it as HE in Figure~\ref{f:flow}(b)) is used for polynomial operations in all the steps other than non-polynomial operations, e.g., $SoftMax$. Instead, GC is used for non-polynomial operations. 
\vspace{-0.1in}
\begin{figure}[ht!]
\centering
\includegraphics[width=3.5in]{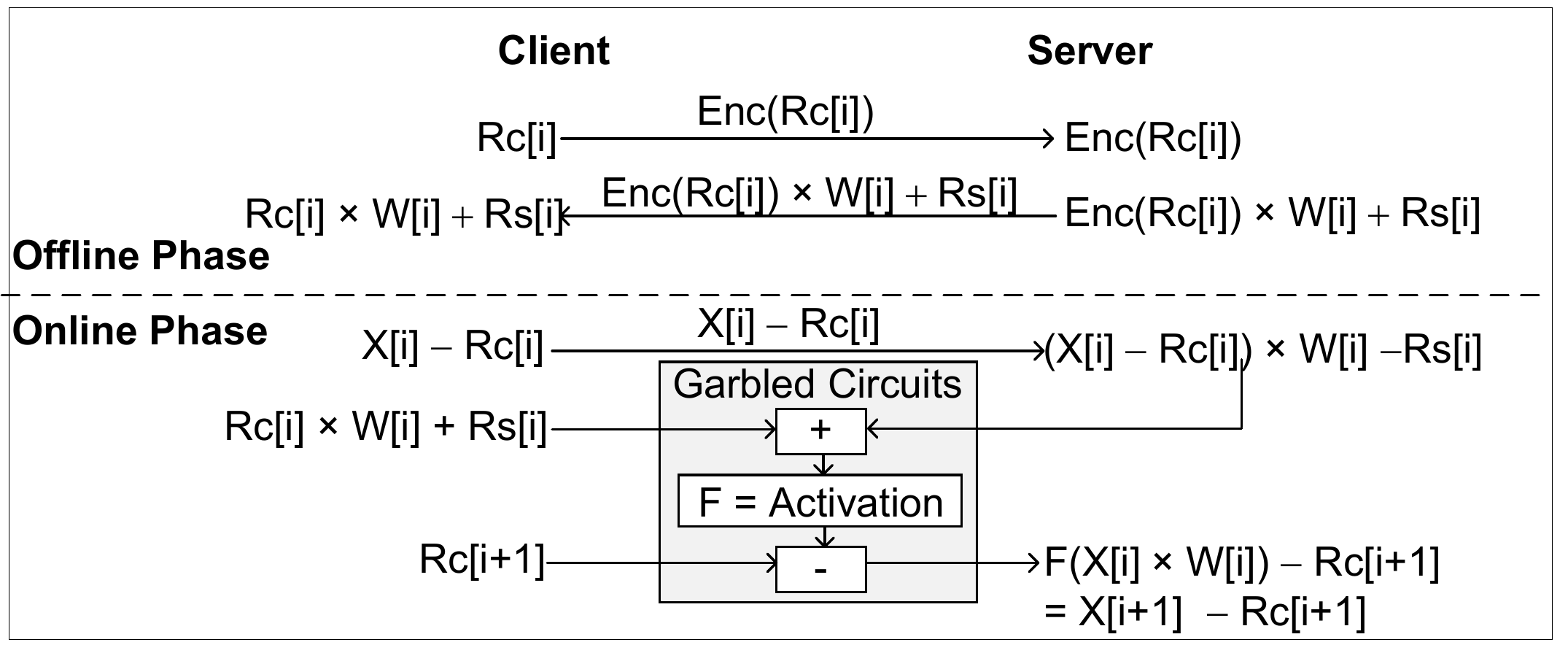}
\caption{Our HGS protocol for Transformer's attention operations.}
\label{f:HGS-module}
\end{figure}
\vspace{-0.15in}
\subsection{Primer-F Construction}
\label{s:Primer-F}

Primer-base significantly improves the accuracy of FHE-based methods like THE-X~\cite{chen:ACL2022} and reduces the latency of MPC-based methods, like GCFormer~\cite{Deepsecure:dac18}.  However, Primer-base still suffers from enormous online latency. This motivates us to propose techniques like the HGS and FHGS protocol, denoted by Primer-F, to offload the online computation to the offline phase where computations can be computed before inference.  

\textbf{The HGS protocol.}
Figure~\ref{f:HGS-module} shows the mixed HGS protocol. The offline phase in the HGS protocol is used to prepare data for the subsequent online phase. For the $i$-th layer of a Transformer model, a client first samples a random matrix $Rc[i]$ that is required to have the same size with private input $X[i]$, and then submits the ciphertext Enc($Rc[i]$) to the server for the subsequent multiplication between Enc($Rc[i]$) and the $i$-th layer weights $W[i]$. A random matrix $Rs[i]$ is generated by the server and $Enc(Rs[i]+Rc[i]\times W[i] )$ is sent back to the client. The client performs decryption to get $Rs[i]+Rc[i]\times W[i]$. $Rs[i]+Rc[i]\times W[i]$ held by the client, and $Rs[i]$ held by the server are secret shares of $Rc[i]\times W[i]$.
Meanwhile, the offline phase, e.g. garbling, of GC is performed. During the online phase, the difference of $X[i]$ and $Rc[i]$, instead of $X[i]$, is sent to the server. The computation of $(X[i]-Rc[i])\times W[i]-Rs[i]$ and previous offline computation make the client and server have the additive secret shares of $X[i]\times W[i]$. In this way, the heavy encrypted HE operations of privacy-preserving  matrix multiplication of $X[i]\times W[i]$ is calculated offline, and the online overhead is almost removed since only unencrypted computations exist. Then GC is used to perform the subsequent mapping function $F$, e.g., $ReLU$ activation. Specifically, the garbled Boolean $X[i]\times W[i]$ is derived by the modular sum of secret shares of $X[i]\times W[i]$, then the Boolean circuits of mapping function $F$ are calculated. Finally, a modular subtraction between function $F$'s result and a new random matrix $Rc[i+1]$ is performed to generate secret shares of function $F$'s result. A modular operation circuit is implemented by an adder and a multiplexer~\cite{GAZELLE:USENIX18,Delphi:usenix2020}.

 We encapsulate HGS protocol shown in Figure~\ref{f:HGS-module} into a module that takes random matrices $Rc[i]$, $Rc[i+1]$, $i$-th layer input $X[i]$, weight matrix $W[i]$ as inputs, and generates $X[i+1]-Rc[i+1]=F(X[i]\times W[i])-Rc[i+1]$. Here $F()$ function can be an identity function or an activation function. The $i$-th layer input $X[i]$ can be removed if the server holds $X[i]-Rc[i]$. $W[0]=W_E, W[2]=\sigma =1$, and $\lambda$ is added to $X[1]$ instead of multiplication. As Figure~\ref{f:flow}(c) shows, steps in the Transformer including \ding{182}  \ding{183}, \ding{184},  \ding{186}, and the other FC computations can be performed by the HGS protocol; however, steps \ding{185} and \ding{187} cannot be directly constructed by the additive HGS protocol since HGS only supports additive computations including ciphertext additions and ciphertext-plaintext multiplication. This is because the HGS protocol that depends on an additive HE scheme is only sufficient for modules where weights are always not encrypted. Therefore, HGS cannot transform ciphertext-ciphertext multiplications in steps \ding{185} and \ding{187} on secret shares into the offline phase. 

\vspace{-0.15in}
\begin{figure}[ht!]
\centering
\includegraphics[width=3.5in]{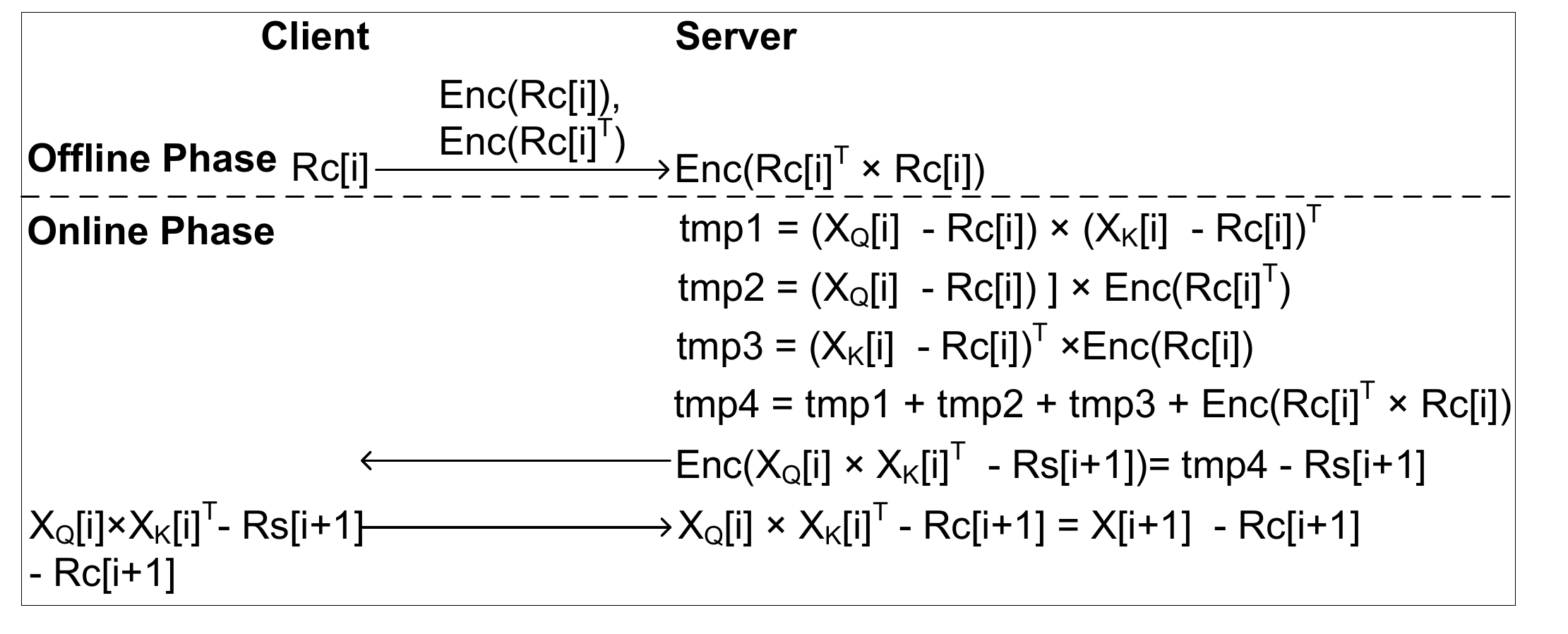}
\caption{Our Fully HGS (FHGS) protocol for attention operations.}
\vspace{-0.1in}
\label{f:FHGS-module}
\end{figure}

\textbf{The Fully HGS (FHGS) protocol for ciphertext-ciphertext operations.}
The step~\ding{185} of attention in Transformer is different from  steps~\ding{183} through \ding{184} where weights are not encrypted and only inputs are encrypted so that private inference is a type of ciphertext-plaintext operations. In step~\ding{185}, however, all query, key, and value matrices are encrypted. The Attention operations require ciphertext-ciphertext operations. Ciphertext-ciphertext operations are not only more expensive than ciphertext-plaintext operations but also cannot directly use  HGS method, thus we propose the FHGS protocol to solve this problem.

Inspired by Beaver's Triple method~\cite{beaver1995}, we propose a Fully HGS (FHGS) protocol to empower the prior additive HGS to efficiently support ciphertext-ciphertext operations such as $X_Q \times X_K^T$  in Transformer models. Figure~\ref{f:FHGS-module} shows our FHGS protocol for step~\ding{185} $X_Q[i]\times X_K[i]^T$. Since $X_Q[i]$ and $X_K[i]^T$ are both ciphertexts, additive HGS cannot offload $X_Q[i]\times X_K[i]^T$ operations. FHGS pre-computes encrypted triples including $Enc(Rc[i])$, $Enc(Rc[i]^T)$, and $Enc(Rc[i]^T\times Rc[i])$ for the usage of the subsequent online process. During the online phase in FHGS, the server has access to $X_Q[i]-Rc[i]$ and $(X_K[i]-Rc[i])^T$ although $X_Q[i]$ and $X_K[i]^T$ are not seen by the server. 
So an important intermediate result $tmp1=(X_Q[i]-Rc[i])\times(X_K[i]-Rc[i])^T$ can be derived. The key idea to obtain our target $tmp4$, a ciphertext of $X_Q[i]\times X_K[i]^T$, is that it can be calculated by subtracting three entries from $tmp1$, where this subtraction can be done by $tmp4=tmp1+tmp2+tmp3+Enc(Rc[i]^T\times Rc[i])$.
In order to preserve the privacy of $X_Q[i]\times X_K[i]^T$, the server transmits its additive secret sharing ciphertext $Enc(X_Q[i]\times X_K[i]^T-Rs[i+1])$, instead of $Enc(X_Q[i]\times X_K[i]^T)$, to the client who can decrypt $Enc(X_Q[i]\times X_K[i]^T-Rs[i+1])$ and obtain $X_Q[i]\times X_K[i]^T-Rs[i+1]$. In this way, the client and the server acquire additive secret shares of $X_Q[i]\times X_K[i]^T$. Optionally, the client can further share $X_Q[i]\times X_K[i]^T-Rs[i+1]-Rc[i+1]$ with the server to obtain  new secret shares of $X_Q[i]\times X_K[i]^T$. At last, the FHGS protocol is enclosed into a module that takes random matrices $Rc[i]$, $Rc[i+1]$, $Rs[i+1]$, $X_Q[i]-Rc[i]$, and $(X_K[i]-Rc[i])^T$ as inputs, and outputs the secret shares of $X_Q[i]\times X_K[i]^T$. 

\textbf{Privacy analysis.}
The $X_Q[i]$, $X_K[i]$, $X_Q[i]\times X_K[i]^T$ are confidential to both the client and the server, which ensures our FHGS protocol is privacy-preserving. The server that has no access to HE private key cannot decrypt ciphertexts including $Enc(Rc[i]^T\times Rc[i])$, $tmp1$, $tmp2$, $tmp3$, $tmp4$, and $tmp4-Rs[i+1]$. Only secret shares of $X_Q[i]$, $X_K[i]$, $X_Q[i]\times X_K[i]^T$ can be accessed by client and server. Also, FHGS completely offloads complex and expensive ciphertext-ciphertext operations from the online phase into the offline phase since $Rc[i]$ and $Rc[i]^T$ are pre-sampled and their product can be calculated in advance, which enables a additive HE scheme to efficiently perform privacy-preserving ciphertext-ciphertext Transformer operations.  

\subsection{Primer-FC by Combined FHGS (CHGS)}
We further reduce the computational and communicational overhead of previous technologies by a combined FHGS (CHGS) method that can combine adjacent HGS layers. The CHGS processes multiple stacked operations using a single calculation, and most HE-based operations are moved to the offline phase. As Figure~\ref{f:flow}(d) shows, our CHGS module removes its previous HGS operations by incorporating three HGS modules into the adjacent FHGS module. The key idea is that the combined target is $(X[i]\times W_E +\lambda)\times W_Q \times [(X[i]\times W_E +\lambda)\times W_K]^T=X_Q[i]\times X_K[i]^T$ which can be derived from $tmp1=((X[i]-Rc[i])\times W_E +\lambda)\times W_Q \times [((X[i]-Rc[i])\times W_E +\lambda)\times W_K]^T$ by $tmp1-tmp2+tmp3+tmp4-tmp5+tmp6-tmp7=result$. The combined weight $W_M$, $tmp6$, and $tmp7$ can be calculated in the offline phase. The server sends $result$-$Rs[i+1]$ to the client so that the client obtains the decryption of $result$-$Rs[i+1]$ and the server has $Rs[i+1]$. The client can also subtract the decryption of $result$-$Rs[i+1]$ with $Rc[i+1]$ to construct a new secret sharing. Using CHGS, 4-time interactions in Figure~\ref{f:flow}(d) can be reduced into 1-time interaction. The improvement details of CHGS are discussed in the following results section.

\begin{figure}[ht!]
\centering
\vspace{-0.1in}
\includegraphics[width=3.4in]{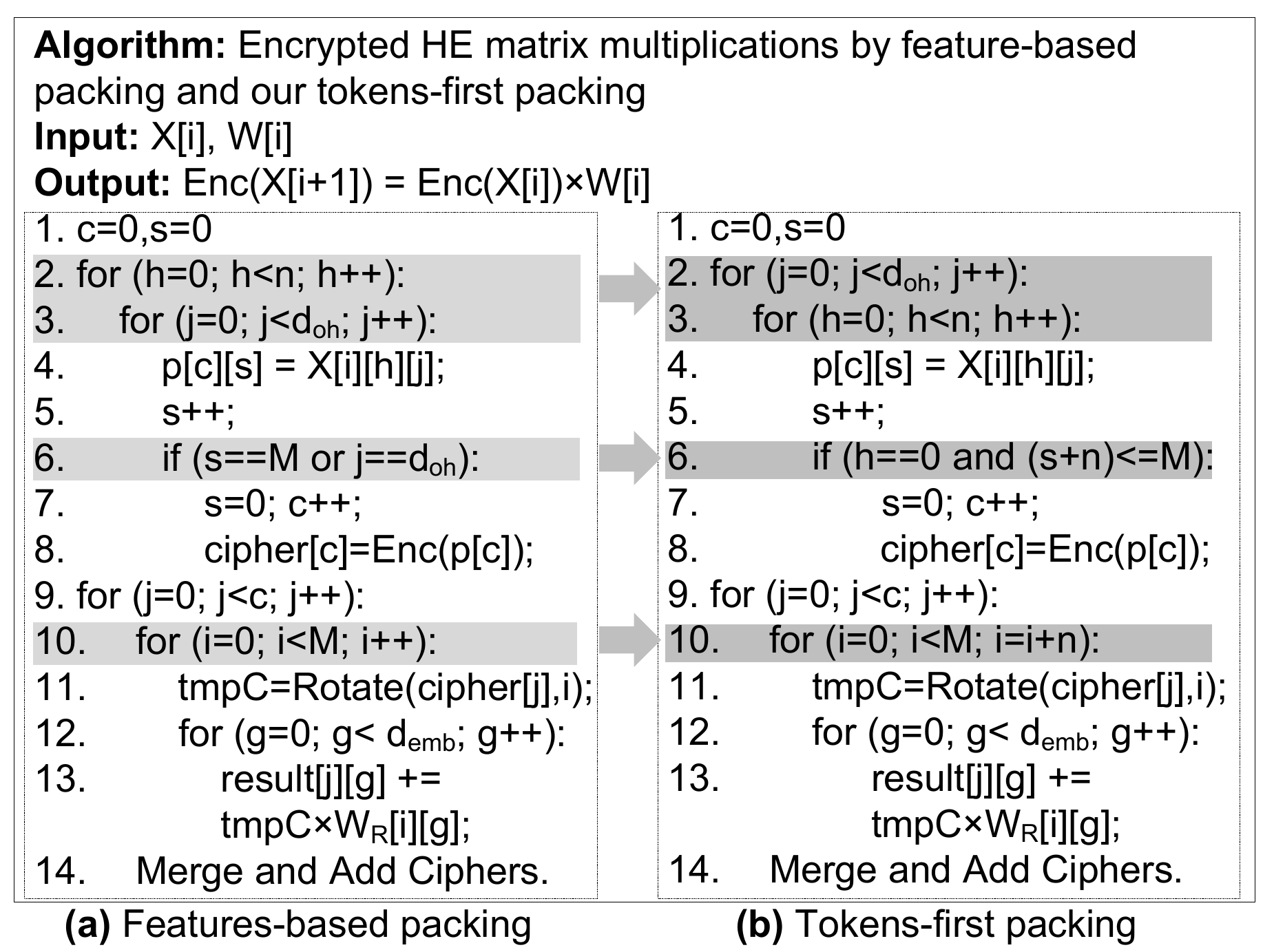}
\caption{The comparison of features-based packing and our tokens-first packing. 
}
\vspace{-0.25in}
\label{f:PACK-CHGS}
\end{figure}

\subsection{Tokens-first Packing, i.e., Primer-FPC Construction}
Embedding is used to compress a large and sparse one-hot vector into a small and dense vector. How to efficiently support high-dimension ($>30K$) matrix multiplication is not studied. Each word (token) will be a vector of size $30522$ which is larger than the ciphertext slot numbers. For multiple words in a sentence, how to pack these words into ciphertext is a new challenge. We propose tokens-first packing to tackle this challenge, instead of prior feature-based ciphertext packing used in~\cite{Brutzkus:ICML2019,GAZELLE:USENIX18, Delphi:usenix2020, Roshan:PLDI20}. In feature-based ciphertext batching,  multiple features (e.g. pixels in an input image) are batched into the same ciphertext. However, we found that directly applying the feature-based packing method on Transformer-based NLP models introduces massive FHE rotations. 

We propose a tokens-first packing method instead of the prior features-based packing method to reduce the homomorphic rotations in Primer-FC. Figure~\ref{f:PACK-CHGS}(a) depicts the pseudo-code of encrypted matrix multiplication based on feature-base packing between $X[i]$, i.e., $X[i][0:n][0:d_{oh}]$ and $W[i]$, i.e., $W[i][0:d_{oh}][0:d_{emb}]$, where $n$, $d_{oh}$, and $d_{emb}$ are input tokens number, one-hot dimension of a token, and embedding dimension, respectively. 
Here $X[i][0:n][0:d_{oh}]$ represents the shape of matrix $X[i]$. The result of matrix multiplication is $X[i+1]$ with a size of $X[i+1][n][d_{emb}]$. Lines 2 through 8 are used to pack the input matrix $X[i][0:n][0:d_{oh}]$ into $c$ plaintexts $p[0:c]$ and encrypt them into $c$ ciphertexts via the encryption function $Enc()$. Each plaintext $p$ has $M$ slots so it can hold $M$ entries. In the features-based packing, the one-hot features $X[i][h][0:j]$ of a token $h$ are first placed into plaintexts, then the features of the next token $h+1$ are packed until all tokens' features are packed and encrypted. The features-based packing requires $c\times M$ $Rotate$ operations since each ciphertext with $M$ features needs $M$ rotations shown in line 9 $\sim$ line 14. One key observation is that features from different tokens are independent and they are not required to accumulate in the matrix multiplication, which motivates us to propose tokens-first packing to batch tokens as much as possible into the same ciphertext. As Figure~\ref{f:PACK-CHGS}(b) shows, the lines of 2, 3, 6, and 10 of features-based packing are replaced so that the  $j$-th feature $X[i][0:n][j]$ of all $n$ tokens are packed into a ciphertext, then the $j+1$-th feature will be packed. Using this tokens-first packing, one ciphertext only has $\sim\frac{M}{n}$ features so that one ciphertext only requires $\sim\frac{M}{n}$ $Rotate$ operations. Considering both features-based packing and our tokens-first packing have similar ciphertext numbers $c$, our tokens-first packing reduces $c\times (M-\frac{M}{n})$ rotations.

\vspace{-0.1in}
\begin{table}[htbp!]
\centering
\small
\setlength{\tabcolsep}{4pt}
\caption{Comparison of our method and other works on private BERT-base inference. "/" means non-applicable. }
\begin{tabular}{l|cccc}
\toprule
\multirow{1}{*}{Scheme}   &   Offline(s)    & Online(s)      &  Total(s) & Acc.(\%) \\
\midrule 
THE-X      & $/$  & $4.7K$ & $4.7K$ & $77.3\%$ \\
GCFormer    &$7.5K$  & $9.8K$ & $17.3K$ & $85.1\%$\\
Primer-F     &$6.5K$ & $0.04K$& $6.54K$ & $84.6\%$ \\
Primer-FPC (Primer)       &$0.4K$ & $0.04K$ & $0.44K$ & $84.6\%$\\






\bottomrule
\end{tabular}
\label{t:overall_comparisons}
\end{table}

\begin{table*}[htbp!]
\centering
\scriptsize
\setlength{\tabcolsep}{4pt}
\caption{A performance ablation study of proposed techniques. Primer-base is constructed by a hybrid cryptographic protocol, where FHE is used for polynomial operations and MPC is for non-polynomial operations. Primer-F, Primer-FP, and Primer-FPC apply FHGS protocols, tokens-first packing, and CHGS techniques, in a cascade manner. The offline and online latency (seconds) of each step in BERT-base are listed. Acc. means accuracy on the MNLI-m dataset. "/" means non-applicable.}
\begin{tabular}{l|cc|cc|cc|cc|cc|cc|cc|c}
\toprule
Scheme   & \multicolumn{2}{c}{\ding{182}\ding{183}  $Embed$}  & \multicolumn{2}{c}{\ding{184}  $QKV$} & \multicolumn{2}{c}{\ding{185} $Q\times K$} & \multicolumn{2}{c}{\ding{186} $SoftMax$} & \multicolumn{2}{c}{\ding{187} $Atten. Value$} & \multicolumn{2}{c}{ Others}  & \multicolumn{2}{|c|}{Total} & Acc.(\%)\\\hline
& offline  & online & offline  & online  & offline  & online & offline  & online & offline  & online & offline  & online & offline  &online
& \\ \midrule

Primer-base     & $/$ & $3094.4$     &   $/$     & $190.6$         &   $/$          & $25$         & $0.01$      & $16.4$  & $0.8$  & $2.3$    & $/$      &  $3224.5$  & $0.81$  & $6553.2$ & $84.6$\\
+FHGS (Primer-F)     & $3094.4$ & $0.8$      & $190.6$        & $3.9$         & $14$             & $9.9$         & $0.01$      & $16.4$  & $0.8$  & $2.3$    & $3224.5$      & $7.9$    &$6524.3$  & $41.2$ & $84.6$\\

+Pack (Primer-FP)    & $134.5$  & $0.8$       & $39$       & $3.9$         & $10.5$            & $7.7$         & $0.01$      & $16.4$        & $0.8$  & $2.3$    & $220.4$      & $7.9$     & $405.2$  & $39$    & $84.6$\\

+CHGS (Primer-FPC)     & $0$  & $0$       & $0$       & $0$       &  $178.2$    & $11.6$ & $0.01$      & $14.8$  & $0.8$  & $2.3$    &  $220.4$     & $6.7$     & $399.4$ & $35.4$ & $84.6$\\






\bottomrule
\end{tabular}
\label{t:overall_ablation_study}
\vspace{-0.1in}
\end{table*}

\vspace{-0.2in}
\begin{table*}[htbp!]
\centering
\scriptsize
\setlength{\tabcolsep}{5pt}
\caption{The performance of Primer over various BERT models on multiple datasets.}
\vspace{-0.1in}
\begin{tabular}{l|cccc|ccccc|cc|c|c}
\toprule
Model   & \multicolumn{4}{c|}{Hyper-parameters}  & \multicolumn{5}{c|}{GLUE and SQuAD Accuracy (\%)} & \multicolumn{2}{c|}{Latency(s)}& Throughput & \multicolumn{1}{c}{Message}\\\hline
& $N$  & $d_{emb}$ & $H$ &  $n$  & MNLI-m  & MRPC  & SST-2  & SQuAD1 & SQuAD2 & offline  &online & tokens/s & GB\\ \midrule

BERT-tiny  & $3$  & $768$ & $12$  & $30$ & $77.6$ & $79.3$   & $88.2$          & $86.2$  & $76.6$  & $318.5$    & $10.6$ & $2.83$  & $0.9$ \\
BERT-small & $6$  & $768$ & $12$  & $30$ & $81.6$ & $84.5$  & $91.1$       & $88.5$ & $78.2$  &  $345.2$    & $18.9$     & $1.59$ & $1.8$ \\
BERT-base  & $12$ & $768$ & $12$  & $30$ & $84.6$ & $86.3$ &$92.5$             & $90.7$  & $80.3$  & $399.4$   & $35.4$    & $0.85$ & $3.6$\\
BERT-medium& $12$  & $1024$ & $16$ & $30$ & $85.4$ & $86.3$ & $93.1$     & $92.2$  & $81.6$  & $452.8$    & $45.1$    &$0.67$& $3.9$\\
BERT-large  & $24$  & $1024$ & $16$ & $30$ & $86.6$ & $87.6$ & $93.5$     & $93.1$ & $82.9$     & $586.4$  &$91.6$     &$0.33$ & $7.9$\\
\bottomrule
\end{tabular}
\label{t:overall_comp_bert_all1}
\vspace{-0.15in}
\end{table*}

\section{Experimental Methodology}
\label{s:exp}
\textbf{System setup and security analysis.} We run the privacy-preserving Transformer experiments on two instances that are equipped with an Intel Xeon E7-4850 CPU and 128 GB DRAM, and each instance was provided with 4 threads. In our current system setup, the average network delay between these two instances is 2.3 ms and the bandwidth is about 100 MB/s. 
The layer-wise PAHE used in Primer is implemented by SEAL~\cite{sealcrypto} libraries where only additive HE operations and rotations are used and ciphertext-ciphertext multiplications are not required. We adopt an extension version of JustGarble tool~\cite{Mihir:Justgarble} used in~\cite{GAZELLE:USENIX18} to implement GC-based operations, including additions of secret sharings and activation functions. The HE parameters and GC settings are selected to provide 128-bit security level. The inputs and weights use 15-bit fix-point representation and the intermediate results are truncated into 15 bits to avoid overflow. The training, fine-tuning, and testing of the Transformer on plaintext was implemented in Python on 4 NVIDIA Tesla V100 GPUs. 


\textbf{Transformer architecture and NLP datasets.} We evaluated Primer on five discriminative NLP models shown in Table~\ref{t:overall_comp_bert_all1}: BERT-Tiny, BERT-small, BERT-base, BERT-medium, BERT-large. The hyper-parameters of these models are listed in Table~\ref{t:overall_comp_bert_all1}. For example, the BERT-tiny model has $N=3$ blocks, $d_{emb}=768$ embedding dimensions, $H=12$ attention heads, and $n=30$ input tokens.  Datasets for five BERT tasks are SQuAD1~\cite{Pranav:SQuAD}, SQuAD2~\cite{lee:etal-2020:squad2}, and MNLI-m, MRPC, SST-2 from GLUE benchmarks~\cite{wang:emnlp2018:glue}.

\section{Results and Analysis}
\label{s:res}
\textbf{Comparison with Prior Works.} We compare our primer with prior works on private BERT-base inference for MNLI-m dataset in table \ref{t:overall_comparisons}. Prior work, e.g., THE-X~\cite{chen:ACL2022} that only uses FHE for private inference only achieves $77.3\%$ accuracy with $4.7k$ seconds latency due to polynomial approximation and expensive FHE operations. We use prior GC-based work~\cite{Deepsecure:dac18} to implement a GCFormer. It achieves an accurate performance, i.e., $85.1\%$ accuracy, but it takes a larger latency than THE-X. Our Primer-F significantly improves
the accuracy of THE-X, e.g., $7.3\%$ accuracy increase, and
    reduces the latency of GCFormer. To reduce the large offline latency of Primer-F, we further propose Primer-FPC, i.e., Primer, with tokens-first packing and combined FHGS. Our primer only takes $\sim 0.4k$-second latency, thus achieving a $\sim 16\times$ latency reduction.

\textbf{Ablation Study.}
Table~\ref{t:overall_ablation_study} describes the performance breakdown and the ablation effects of proposed techniques using BERT-base model with $n=30$ on MNLI-m dataset. The Primer-base implemented by FHE and MPC protocols requires $\sim$ 6553 seconds latency to perform one inference on a sentence in MNLI-m dataset and achieves $84.6\%$ accuracy. We further propose FHGS, denoted by Primer-F to  Offload computations into the offline phase from the online phase, which significantly shrinks the offline latency from $\sim 6553$ seconds to $\sim 41$ seconds, introducing almost $160\times$ latency reduction. Primer-FP is proposed to reduce the latency of embedding layers and the following layers that include HE operations. It further decreases $5.3\%$ online latency over Primer-F and has $16\times$ offline latency reduction. Primer-FPC has the similar offline latency and accuracy with Primer-topk, but further reduces the online latency by $9.2\%$.  Table~\ref{t:overall_ablation_study} shows Primer (Primer-FPC) achieves competitive NLP accuracy and reduces the online and offline inference latency by $90.6\% \sim 97.5\%$ over Primer-base.

\vspace{0.1in}
\textbf{Results on Different Models.} Table~\ref{t:overall_comp_bert_all1} studies the effects of different BERT models and datasets on Primer. Privacy-preserving BERT-tiny with only 3 Transformer blocks achieves average $81.7\%$ accuracy on three GLUE datasets including MNLI-m , MRPC, SST-2, and average $81.4\%$ F1 test accuracy on SQuAD1 and SQuAD2. Primer with BERT-tiny requires 10.6 seconds to perform an inference or classification on a sequence with $30$ tokens, thereby attaining a throughput of 2.83 tokens per second. Also, the communicational message size between clients and a server is $\sim 0.9$ GB. Primer with BERT-small or BERT-base achieves $2.4\% \sim 7.5\%$ higher accuracy by adding more Transformer blocks, but increases latency by $78.3\% \sim 230\%$. Also, Primer with BERT-small and BERT-base require more than 2$\times$ and 3 $\times$ message size, respectively, than BERT-tiny based Primer.  BERT-medium with larger embedding dimension and BERT-large with 24 block numbers are also supported by Primer, and they take 45.1 seconds and 91.6 seconds to perform an inference on a sentence with $30$ tokens.  BERT-large achieves state-of-the-art accuracy on GLUE and SQuAD benchmarks.


\section{Conclusion}
In this paper, we present Primer to enable a fast and accurate privacy-preserving Transformer for NLP tasks. First, a na\"ive version of our Primer, called Primer-base, is constructed by a hybrid interactive cryptographic protocol. Secondly, we propose tokens-first packing instead of prior features-first packing to reduce the offline and online overhead brought by HE. Finally, we demonstrate that multiple secret sharing layers in the Transformer can be combined to reduce the latency. Primer establishes a solid baseline and shed the light on private Transformer inference over encrypted data. 

\section*{Acknowledgment}
This work was supported in part by NSF awards CCF-1908992, CCF-1909509, and CCF-2105972. 

\bibliographystyle{short}
\bibliography{hebib}

\begin{thebibliography}{10}
\newcommand{\enquote}[1]{``#1''}
\def\url#1{}
\providecommand{\urlprefix}{}

\bibitem{beaver1995}
D.~Beaver, \enquote{Precomputing oblivious transfer,} in \emph{Annual
  International Cryptology Conference}, pages 97--109, Springer, 1995.

\bibitem{Mihir:Justgarble}
M.~Bellare, \emph{et~al.}, \enquote{Efficient Garbling from a Fixed-Key
  Blockcipher,} Cryptology ePrint Archive, Report 2013/426, 2013.

\bibitem{Brutzkus:ICML2019}
A.~Brutzkus, \emph{et~al.}, \enquote{Low latency privacy preserving inference,}
  in \emph{International Conference on Machine Learning}, pages 812--821, 2019.

\bibitem{chen:ACL2022}
T.~Chen, \emph{et~al.}, \enquote{{THE}-{X}: Privacy-Preserving Transformer
  Inference with Homomorphic Encryption,} in \emph{Findings of the Association
  for Computational Linguistics: ACL 2022}, pages 3510--3520, Association for
  Computational Linguistics, Dublin, Ireland, May 2022.

\bibitem{Roshan:PLDI20}
R.~Dathathri, \emph{et~al.}, \enquote{EVA: An Encrypted Vector Arithmetic
  Language and Compiler for Efficient Homomorphic Computation,} in
  \emph{Proceedings of the 41st ACM SIGPLAN Conference on Programming Language
  Design and Implementation}, PLDI 2020, page 546–561, Association for
  Computing Machinery, New York, NY, USA, 2020.

\bibitem{Devlin:ACL2019:BERT}
J.~Devlin, \emph{et~al.}, \enquote{{BERT}: Pre-training of Deep Bidirectional
  Transformers for Language Understanding,} in \emph{Proceedings of the 2019
  Conference of the North {A}merican Chapter of the Association for
  Computational Linguistics: Human Language Technologies, Volume 1 (Long and
  Short Papers)}, pages 4171--4186, Association for Computational Linguistics,
  Minneapolis, Minnesota, June 2019.

\bibitem{feng:2020:cryptogru}
B.~Feng, \emph{et~al.}, \enquote{CryptoGRU: Low Latency Privacy-Preserving Text
  Analysis With GRU,} \emph{arXiv preprint arXiv:2010.11796}, 2020.

\bibitem{FHE}
C.~Gentry, \enquote{Fully homomorphic encryption using ideal lattices,} in
  \emph{Proceedings of the forty-first annual ACM symposium on Theory of
  computing}, pages 169--178, 2009.

\bibitem{GMW:1987:SS}
O.~Goldreich, \emph{et~al.}, \enquote{How to Play ANY Mental Game,} in
  \emph{Proceedings of the Nineteenth Annual ACM Symposium on Theory of
  Computing}, STOC '87, page 218–229, Association for Computing Machinery,
  New York, NY, USA, 1987.

\bibitem{GAZELLE:USENIX18}
C.~Juvekar \emph{et~al.}, \enquote{{GAZELLE: A Low Latency Framework for Secure
  Neural Network Inference},} in \emph{{USENIX Security Symposium}}, 2018.

\bibitem{lee:etal-2020:squad2}
G.~Lee, \emph{et~al.}, \enquote{{SQ}u{AD}2-{CR}: Semi-supervised Annotation for
  Cause and Rationales for Unanswerability in {SQ}u{AD} 2.0,} in
  \emph{Proceedings of the 12th Language Resources and Evaluation Conference},
  European Language Resources Association, Marseille, France, May 2020.

\bibitem{li2022mpcformer}
D.~Li, \emph{et~al.}, \enquote{MPCFormer: fast, performant and private
  Transformer inference with MPC,} \emph{arXiv preprint arXiv:2211.01452},
  2022.

\bibitem{Lou:NIPS2019}
Q.~Lou and L.~Jiang, \enquote{SHE: A Fast and Accurate Deep Neural Network for
  Encrypted Data,} in \emph{Advances in Neural Information Processing Systems},
  pages 10035--10043, 2019.

\bibitem{Lou:NIPS20:AutoPrivacy}
Q.~Lou, \emph{et~al.}, \enquote{AutoPrivacy: Automated Layer-wise Parameter
  Selection for Secure Neural Network Inference,} in \emph{Advances in Neural
  Information Processing Systems}, edited by H.~Larochelle, \emph{et~al.},
  volume~33, pages 8638--8647, Curran Associates, Inc., 2020,
  \urlprefix\url{https://proceedings.neurips.cc/paper/2020/file/6244b2ba957c48bc64582cf2bcec3d04-Paper.pdf}.

\bibitem{Lou:ICLR2021:safenet}
Q.~Lou, \emph{et~al.}, \enquote{SAFENet: A Secure, Accurate and Fast Neural
  Network Inference,} in \emph{International Conference on Learning
  Representations}, 2021.

\bibitem{Delphi:usenix2020}
P.~Mishra, \emph{et~al.}, \enquote{Delphi: A Cryptographic Inference Service
  for Neural Networks,} in \emph{{USENIX Security Symposium}}, {USENIX}
  Association, Boston, MA, August 2020.

\bibitem{radford:2019:GPT2}
A.~Radford, \emph{et~al.}, \enquote{Language models are unsupervised multitask
  learners,} \emph{OpenAI blog}, 1(8):9, 2019.

\bibitem{Pranav:SQuAD}
P.~Rajpurkar, \emph{et~al.}, \enquote{SQuAD: 100, 000+ Questions for Machine
  Comprehension of Text,} \emph{CoRR}, abs/1606.05250, 2016.

\bibitem{Deepsecure:dac18}
B.~D. Rouhani \emph{et~al.}, \enquote{{DeepSecure: Scalable Provably-Secure
  Deep Learning},} in \emph{{ACM/IEEE Design Automation Conference}}, 2018.

\bibitem{sealcrypto}
\enquote{{M}icrosoft {SEAL} (release 4.0),}
  \url{https://github.com/Microsoft/SEAL}, March 2022, microsoft Research,
  Redmond, WA.

\bibitem{Vaswani:NIPS2017:Attention}
A.~Vaswani, \emph{et~al.}, \enquote{Attention is All you Need,} in
  \emph{Advances in Neural Information Processing Systems}, edited by I.~Guyon,
  \emph{et~al.}, volume~30, Curran Associates, Inc., 2017.

\bibitem{wang:emnlp2018:glue}
A.~Wang, \emph{et~al.}, \enquote{{GLUE}: A Multi-Task Benchmark and Analysis
  Platform for Natural Language Understanding,} in \emph{Proceedings of the
  2018 {EMNLP} Workshop {B}lackbox{NLP}: Analyzing and Interpreting Neural
  Networks for {NLP}}, pages 353--355, Association for Computational
  Linguistics, Brussels, Belgium, November 2018.

\bibitem{google:2016:wordpiece}
Y.~Wu, \emph{et~al.}, \enquote{Google's Neural Machine Translation System:
  Bridging the Gap between Human and Machine Translation,} \emph{CoRR},
  abs/1609.08144, 2016.

\end{thebibliography}
\end{document}